\documentclass[journal=jacsat,manuscript=article]{achemso}

\usepackage{graphicx,amssymb,amsmath,epsfig}
\usepackage{color}
\usepackage[latin1]{inputenc}
\usepackage[version=3]{mhchem} 

\author{Davide Donadio}
\affiliation[Max Planck Institute for Polymer Research]{Max Planck Institute for Polymer Research, Ackermannweg 10, D-55128 Mainz, Germany}
\email{donadio@mpip-mainz.mpg.de}
\author{Luca M. Ghiringhelli}
\affiliation[Fritz-Haber-Institut]{Fritz-Haber-Institut, Faradayweg 4--6, D-14195 Berlin-Dahlem, Germany}
\email{ghiringhelli@fhi-berlin.mpg.de}
\author{Luigi Delle Site}
\affiliation[Freie Universit\"{a}t Berlin]{Institute for mathematics, Freie Universit\"{a}t Berlin, Arnimallee, 6, D-14195 Berlin, Germany}
\email{luigi.dellesite@fu-berlin.de}

\title{Autocatalytic and cooperatively-stabilized dissociation of water on a stepped platinum surface}

\begin{document}

\begin{abstract}
Water-metal interfaces are ubiquitous and play a key role in many chemical processes, from catalysis to corrosion. Whereas water adlayers on atomically flat transition metal surfaces have been investigated in depth, little is known about the chemistry of water on stepped surfaces, commonly occurring in realistic situations. Using first-principles simulations we study the adsorption of water on a stepped platinum surface.
We find that water adsorbs preferentially at the step edge, forming linear clusters or chains, stabilized by the cooperative effect of chemical bonds with the substrate and hydrogen bonds. 
In contrast with flat Pt,  at steps water molecules dissociate forming mixed hydroxyl/water structures, through an autocatalytic mechanism promoted by hydrogen bonding. Nuclear quantum effects contribute to stabilize partially dissociated cluster and chains.
Together with the recently demonstrated attitude of water chains adsorbed on stepped Pt surfaces to transfer protons via thermally activated hopping, these findings candidate these systems as viable proton wires.
\end{abstract}

\section{Introduction}

The structure and dynamics of water films on metallic surfaces has received growing attention in the past two decades, with special focus on the details of monolayer and sub-monolayer coverage~\cite{Henderson:2002p3992,Hodgson:2009p3968}.
The interest in understanding water on closed-packed metal surfaces goes beyond the academic curiosity into a complex and challenging problem, or the exploitation of refined surface-science techniques: Water-metal interactions play a central role in many catalytic surface reactions, corrosion processes and electrochemistry, with remarkable impact on fuel cells and photoelectrochemical cells~\cite{Khaselev:1998p4403,Chen:2011p4401}.
By combining experimental surface probes and theoretical calculations, mostly at the level of density functional theory (DFT), the adsorption properties of water on high-symmetry surfaces of several transition metals, such as Cu(110) and (111), Pt(111), Pd(111), Rh(111) and Ru(0001), have been finely characterized~\cite{Hodgson:2009p3968,Carrasco:2012p4557}. Nevertheless in most real-life cases metal surfaces are not atomically flat.
Absorption on high-index surfaces and the effects of extended surface defects, such as steps, terraces or grain boundaries, are however much less studied and understood. 
On surfaces with chain--like features, such as Cu(110), at low coverage, water forms one-dimensional structures, with peculiar molecular arrangement~\cite{Carrasco:2009p3714} and dissociation may be enhanced~\cite{Lee:2008p3974,Andersson:2008p3911}, so to stabilize a partially dissociated wetting layer at full coverage~\cite{Forster:2011p4691}. 
STM observations of the adsorption process on Pt showed that water is initially adsorbed at step edges, forming hydrogen-bonded chains~\cite{Morgenstern:1996p3971}. 
Such water chains were characterized by surface X-ray diffraction experiments~\cite{Nakamura:2009p3970}, showing that water molecules are adsorbed at top site on step edges on Pt(211), and were selectively isolated by thermal desorption~\cite{Picolin:2009p3988}. 
In other words, while at low temperatures two-dimensional water islands are adsorbed on the terraces, for a well defined range of higher temperatures, only the steps are found to be covered by one-dimensional water structures.

Only recently theorists have started to investigate the structure of water on stepped metallic surfaces~\cite{Arnadottir:2010p4110,Lin:2012p4402}. An exploratory theoretical study has demonstrated the possibility of proton transfer in water chains, adsorbed at step edges of metal surfaces~\cite{Scipioni:2011p3934}, leading to the more general question about the protonation state of one-dimensional water structures.
The possibility of stabilizing geometrically-controlled conducting water chains represents an intriguing technological possibility that can be exploited in several microelectronic applications, like electrocatalysts, conductometric gas-phase sensors, batteries, fuel cells and photovoltaics.

Water bilayers adsorb intact on Pt (111), forming $\sqrt{37}\times\sqrt{37}R25.3$ or a $\sqrt{39}\times\sqrt{39}R16.2$ regular pattern, depending on the coverage~\cite{Glebov:1997p4063,Haq:2002p4074}. On the other hand, mixed OH/water monolayers with $\sqrt{3}\times\sqrt{3}R30$ or $3\times 3$ periodicity~\cite{Karlberg:2003p3981} have been obtained by either co-adsorption of water and  oxygen~\cite{Clay:2004p3963} or by electron damage~\cite{Harnett:2003p4065}. Theory predicts that the dissociation reaction of water at Pt(111) becomes thermodynamically favorable for bias potentials above 0.63 eV~\cite{Rossmeisl:2006p4556}.
Both intact and partially dissociated water adlayers are stabilized by the cooperative effect of hydrogen bonding and surface bonding, which takes place between the metal and the lone-pair of water~\cite{DelleSite:2007p24,Schiros:2010p3991}. 
A recent  study on the formation of hydroxyl on Pt in the presence of co-adsorbed oxygen~\cite{Lew:2011p4112} showed that OH$^-$ is more favorably bonded at step sites than at terraces and suggested that endothermic dissociation of H$_2$O to form mixed H$_2$O-OH coverage layers is initiated at the steps. Reflection adsorption IR spectroscopy and X-ray spectroscopy (NEXAFS) indicate that water molecules adsorbed at the step edge of Pt(533) and Pt(211) are strongly hydrogen bonded but cannot exclude partial dissociation~\cite{Grecea:2004p4363,Endo:2012}. 

In this context, theoretical/computational methods represent a predictive tool for zooming into the atomistic process of adsorption and dissociation and provide a pilot path for further experimental investigations. 
Here, by means of DFT calculations, we investigate the combined effects of  increased reactivity and reduced dimensionality on adsorption and dissociation of water at the edge of a step of the Pt(221) surface. 
We analyze the adsorption energy and the dissociation energy and barrier for clusters of increasing number of water molecules and of periodic water wires adsorbed on the step edge of a Pt(221). The estimate of the dissociation energies is refined by including the effects due to the quantum nature of the nuclei. This is done by calculating the zero point energy (ZPE) correction via the evaluation of the harmonic vibrational frequencies. 
We finally interpret the nature of the cooperative effects that stabilize the dissociated chains in terms of electronic density displacements. We show that the cooperative stabilization effect is rather local, i.e. its electronic characterization does not change for dissociation ratios smaller than 1:3. 

On the basis of this analysis we are able to claim that cooperative hydrogen bonding increases the adsorption energy of intact water at the step edge, but tends to de-stabilize the covalent O-H bond, favoring partial dissociation even in absence of co-adsorbed oxygen. Energetically, the optimal ratio of dissociated molecules is 1:4, when the quantum nature of the nuclei is taken into account. These results show that the step edge of Pt(221) is a favorable substrate to enhance water dissociation because of its structure and electronic reactivity.

\section{Methods}
We consider hydrogen bonded clusters and periodic chains of water molecules adsorbed on a stepped Pt (221) surface in a periodic supercell, as depicted for example in Fig.~\ref{Fig:clusters}a. We computed the optimal structures and the energetics of adsorption and dissociation using DFT in the generalized gradient approximation, using the functional by Perdew, Burke and Ernzerhof (PBE)~\cite{Perdew:1996p4021}. Ultrasoft pseudopotentials are used to restrict the calculation to valence electrons~\cite{VANDERBILT:1990p4058} and the electronic wavefunctions are expanded on a plane-wave basis set with a cutoff of 35 Ry. To test this choice of simulation parameters we have considered the adsorption and dissociation energy of the water monomer and dimer. These quantities varied less than 0.01 eV when we tested the convergence of the PW cutoff up to 55 Ry, and when we repeated the calculation with norm-conserving pseudopotentials and a cutoff of 65 Ry.
The PBE functional provides a very good description of hydrogen bonding in bulk ice~\cite{Hamann:1997}, and is in excellent agreement with gold--standard quantum--chemical calculations as for structure and binding energy of the water dimer~\cite{Santra:2007p4145,Zhang:2011} and of small clusters~\cite{Santra:2008p4064}. It also reproduces well the complex structures ($\sqrt{37}$ and $\sqrt{39}$) of the first wetting layer on Pt(111), yielding very good agreement with STM experiments~\cite{Nie:2010p4551}. 
With our setup the absolute absorption energy of a single water molecule is underestimated, compared to that measured by temperature programmed desorption~\cite{SEXTON:1984p4550} (0.26 vs. 0.44 eV), probably due to the lack of dispersion corrections. However, such deficit is systematic and should not affect relative energy differences. Dispersion forces indeed play a crucial role when the binding energy of a water layer is compared to that of ice, but do not make a significant difference between different adsorbate structures~\cite{Carrasco:2011p026101}. 
We have verified that the calculations of the dissociation energy of the monomer adsorbed on Pt(221) step are consistent using either PBE (0.51 eV) or the nonlocal vdW-DF functional~\cite{Gulans:2009} (0.59 eV).~\footnote{However at the edge of acceptable quantitative estimations, an uncertainty of 0.08 eV do not change qualitatively our main conclusions about partial dissociation. It is also worth noting that the current approaches to include vdW dispersion are not universal enough to provide a systematic improvement in the quantitative predictions.}
In general PBE  performs quite well also with respect to H dissociation barriers. According to Ref.~\cite{Zheng:2007p4545} the mean unsigned error for a set representative of H dissociation systems is 0.1 eV, i.e. below the difference in barrier heights that we find for the monomer and cluster cases.

The Pt(221) surface is represented by 4--layer thick slab with a vacuum layer 18 \AA\ thick. The two bottom layers of the slab are kept fixed to mimic the bulk. We verified that adsorption and dissociation energies change less than 0.02 eV, when 2 further layers are added.
The simulation cell is made of a number of replicas of the 221 unit cell, sufficient to accommodate a given number of water molecules in a cluster or in a periodic chain: for example for the monomer we use three replicas, yielding a square surface cell with edge length of 8.64 \AA. To compute the electronic structure the first Brillouin Zone is sampled using a $4\times 4\times 1$ Monkhorst-Pack mesh shifted in its $xy$ components~\cite{Monkhorst}. Equivalent k-point meshes are used for different geometries. The electronic occupation at the Fermi level was smeared using the Methfessel and Paxton approach~\cite{mp} with a Gaussian spread of 0.27 eV. 
The structures are optimized with a convergence threshold of 5$\cdot 10^{-4}$ atomic units on the largest force component.  With this choice of parameters we estimate an accuracy on adsorption and dissociation energies of 0.02 eV.
To characterize the vibrational properties of the adsorbates and estimate the zero point energy (ZPE) corrections to the total energies we computed the normal modes of the adsorbed species using the frozen-phonon approach, i.e. computing the force constant matrix by finite differences of the forces, upon finite displacement of the atoms ($\delta x=0.01$ \AA). 
Dissociation paths and reaction energies were evaluated using the nudge-elastic-band (NEB) method~\cite{Henkelman:2000p4071}, with either seven or nine replicas of the system along the transition path. 
All the calculations are performed using the Quantum--Espresso package~\cite{Giannozzi:2009p3719}.

\section{Results and Discussion}

The models of water clusters and periodic chains considered in this paper are depicted in Figs~\ref{Fig:clusters} and~\ref{Fig:chains}. In Fig.~\ref{Fig:clusters}  the left panels show the configurations with intact molecules and the right panels the corresponding configurations after deprotonation. 
Water molecules adsorb atop at the step edge atoms, and the lattice spacing between Pt atoms (2.82 \AA) is suitable to accommodate hydrogen-bonded chains or clusters, however in some cases not all the water molecules remain attached to the step.
The adsorption energy is defined as:
\begin{equation}
 E_{ads}=  [ E(N\cdot H_2O\textrm{@}Pt) - N\cdot E(H_2O) - E(Pt) ] /N 
\end{equation}
where $N$ is the number of H$_2$O molecules, and $E(N\cdot H_2O$@$Pt)$, $E(H_2O)$ and $E(Pt)$ are the energies of the metal surface with adsorbed water, of a single water molecule and of the free Pt surface, respectively.
The adsorption energy of water clusters and chain on Pt(221) was formerly computed by DFT with slightly different choices as for the exchange and correlation functional (PW91 vs. PBE in our case) and the simulation setup~\cite{Arnadottir:2010p4110}. When similar configurations are considered, as in the case of the monomer, dimer and zig-zag chain, our results agree very well as for adsorption energies and geometries. 

\begin{figure}[t]
\centerline{\includegraphics[width=8.4cm]{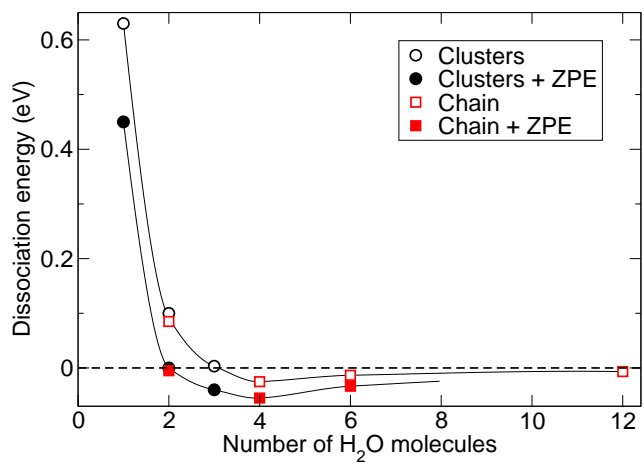}}
\caption{Dissociation energies per water molecule of clusters and periodic chains of $n$ water molecules, with and without zero point energy (ZPE) correction.
The lines are guides to the eyes and connect dissociation energy datasets with and without ZPE corrections, respectively.} 
\label{Fig:energies}
\end{figure}

When one H$_2$O dissociates into a OH$^-$ and a H$^+$, the proton adsorbs favorably at the terrace.
We define the  dissociation energy, $E_{diss}$, as the total energy difference between the system where one water is dissociated into an OH and a proton,  $E((N-1)H_2O+OH+H$@$Pt )$, and the intact water cluster adsorbed on the Pt surface, $E(N\cdot H_2O$@$Pt)$. 
A summary of the calculated binding and dissociation energies is given in Tab.~\ref{Tab:energy}.
We also define the dissociation energy per water molecule $\varepsilon_{diss} = E_{diss} / N $, where $N$ is the number of H$_2$O molecules. We plot $\varepsilon_{diss}$ as a function of the length of the chain in Fig. \ref{Fig:energies}.
We verified that the most favorable adsorption site for the dissociated proton is {\sl fcc} hollow site of the terrace, similar \cite{Michaelides:2001p3671} to the case of Pt(111) surface.
For clusters, protons are preferentially adsorbed at {\sl fcc} hollow sites near the step edge, whereas for periodic water wires  protons are adsorbed at hollow sites in the middle of the terrace. Proton adsorption at the base of the step is instead never advantageous.
Exploratory {\em ab initio} molecular dynamics runs show~\cite{Scipioni:2011p3934} that protons on Pt are highly mobile and diffusion on the terrace may occur over short time scales (ps) even at temperatures as low as 150 K.
    
\subsection{Clusters}

\begin{figure}[htbp]
\centerline{\includegraphics[width=8.4cm]{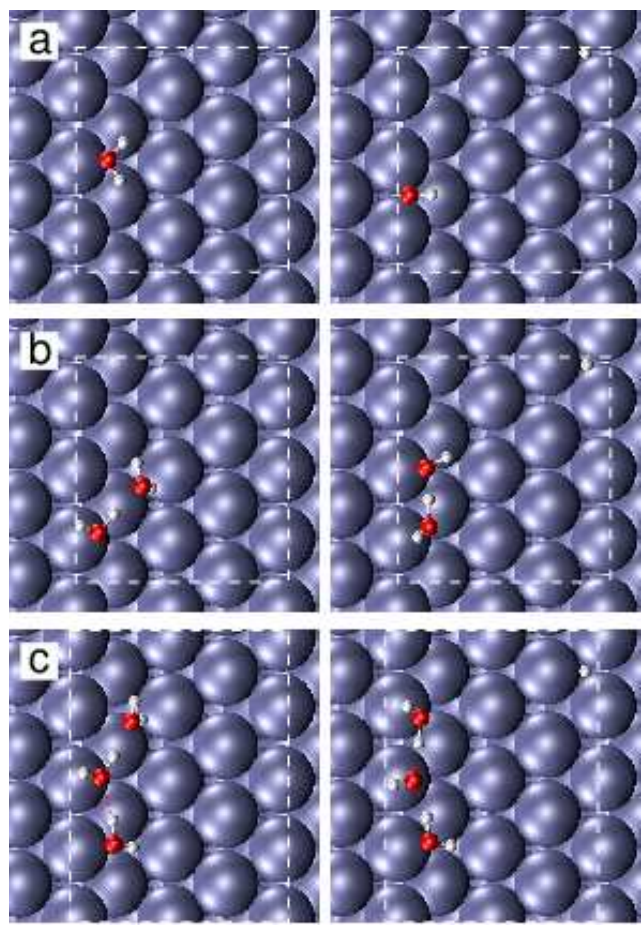}}
\caption{
Structure of water clusters at the step of  Pt(221) surface. Panels (a--c) show the water monomer, dimer and trimer in their intact (left) and dissociated configurations (right). 
The edges of the periodically replicated simulation cells used for our calculations are represented by white dashed lines.
}
\label{Fig:clusters}
\end{figure}

First we analyze the details of the adsorption and dissociation of water monomer, dimer and trimer on the stepped Pt(221) surface. 
The adsorption energies of water clusters at the edge of the step are larger than on Pt(111) by 0.18 eV for the monomer to 0.09 eV for the trimer,
as a consequence of the increased reactivity of the step edge. In the adsorption of intact water dimers and trimers there is a competition between O-Pt  bonding and H-bonding, which results in optimal geometries where one of the water molecules, the hydrogen-bond (HB) acceptor only, detaches from the step edge and forms a weak HB with the lower terrace of the metal, as can be seen in Fig.~\ref{Fig:clusters}b and~\ref{Fig:clusters}c (left panels). 
This configuration facilitates the dissociation of the water molecule detached from the step, lowering the activation energy from 0.8 eV for the monomer to 0.5 eV for dimers and trimers. 
After dissociation, shorter (and then stronger) chemical bonds are formed between all the oxygen atoms and the Pt atoms at the step edge. In the trimer a proton transfer mechanism leads to the stable configuration, where the remaining hydroxyl is the central molecule of the cluster that accepts HBs from the two neighboring water molecules (Fig.~\ref{Fig:clusters}c). 
Whereas in the case of the monomer the energy cost of breaking an O-H bond is only partially compensated by a stronger O-Pt bond, resulting in a large dissociation energy (0.51 eV), the rearrangement of the clusters reduces the dissociation energy to 0.2 eV for the dimer and to nearly zero in the case of the trimer.
Further details on the geometries and dissociation mechanisms of clusters are provided in the Supplementary Information.

\subsection{Periodic water wires}

\begin{figure}[tbp]
\centerline{\includegraphics[width=8.4cm]{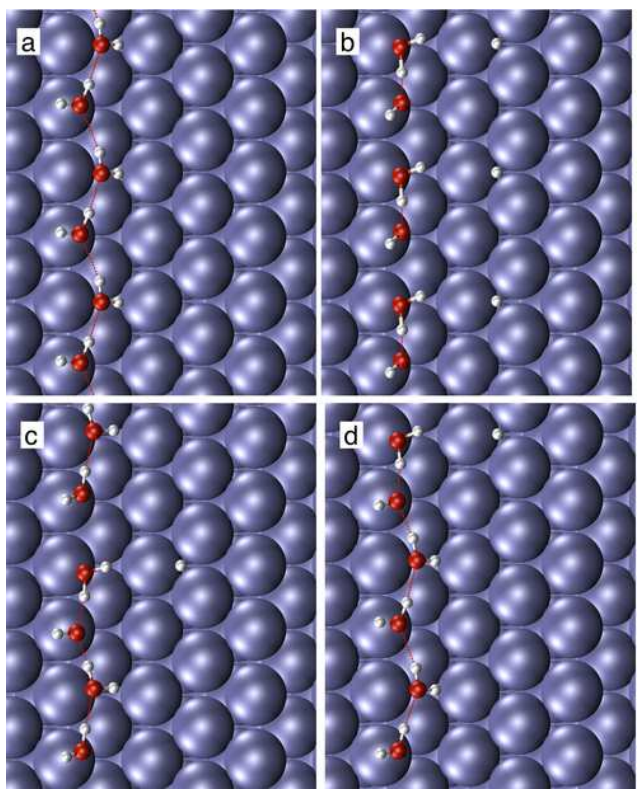}}
\caption{
Structure of water wires adsorbed at the step of  Pt(221) surface. Panel (a) represents a periodic hydrogen--bonded zig-zag chain. Panels (b--d) represent the configuration of the periodic chain when one out of two (b), four (c) or six (d) molecules are dissociated. The graphics shows clearly that the hydrogen-bond chain is interrupted by the rearrangement of the molecules upon deprotonation. 
}
\label{Fig:chains}
\end{figure}

Periodic zig-zag chains of hydrogen-bonded water molecules~\cite{Endo:2012} can easily be accommodated at the step edge of transition metal surfaces. On Pt (221) the chain is formed by alternating short (strong) and long (weak) HBs, with O-O distances of 2.86 and 3.15 \AA, respectively (Fig.~\ref{Fig:chains}a). O-Pt distances are 2.40 and 2.50 \AA. This pattern is modulated by the weak interaction of half of the molecules with the Pt atoms of the lower terrace through a long O-H-Pt bond. 
This complex combination of interactions leads to an adsorption energy of 0.59 eV/H$_2$O, which is slightly higher than that of the trimer. The basic unit cell for the chain includes two water molecules, but relaxing the periodicity up to six molecules has no effect on the equilibrium geometry, nor changes the adsorption energy.

Dissociation is achieved when one of the molecules pointing towards the lower terrace releases a proton. The  rearrangement that follows depends on how many water molecules are dissociated. Here we consider the cases of dissociation of one H$_2$O out of 2, 4, 6 and 12.
When the ratio of dissociated molecules is 1/2 the system arranges into a periodic sequence of  dissociated dimers (Fig.~\ref{Fig:chains}b). The geometry of the periodic dimers and the dissociation energy (0.17 eV) are indeed similar to the case of the dissociated dimer discussed above. 
The hydroxyl groups never act as HB donor, so the chain of HBs is broken, at variance with the bi-dimensional mixed H$_2$O-OH layer. 
If one out of 4 water molecules is deprotonated, the dissociation is followed by proton transfer along the water wire, which rearranges so that the OH$^-$ is bracketed by two intact water molecules and accepts two HBs. Also in this case the HB chain is interrupted. The equilibrium structure after dissociation is a periodic sequence of dissociated hydrogen-bonded tetramers, as shown in Fig.~\ref{Fig:chains}c. The dissociation in this case is clearly exothermic ($E_{diss}=-0.10$ eV).  
Similar dissociation energies ($E_{diss}=-0.08$ eV) and structural relaxation are observed when one out of six and one out of twelve molecules are dissociated (Fig.~\ref{Fig:chains}d). This suggests that the rearrangement is local and increasing the number of degrees of freedom that can rearrange upon dissociation does not produce a significant relaxation.
We can then conclude that the most favorable dissociation rate for periodic chains is one molecule out of four, which corresponds to the minimum in the curve in Fig.~\ref{Fig:energies}.

The activation energy for dissociation is $\sim0.5$ eV, independent on the concentration of dissociated molecules, whereas the proton transfer and rearrangements of the chains are barrier-less. If this value of the barrier is compared to the higher barrier (at least 0.82 eV) to be overcome in the dissociation of a water monomer on a step Pt surface, the function of the non-dissociating water molecules in the chain is that of a catalyst for the proton dissociation. In this sense, we have an autocatalytic dissociation. The barrier of $\sim0.5$ eV is still relatively high at the typical experimental temperature of $\sim155-170 K$ at which water wires can be isolated on steps by desorption of water from the terraces~\cite{Nakamura:2009p3970,Picolin:2009p3988}. Yet, using a simplified expression for the rate, $k = \bar{\nu}~exp(- \Delta E / k_BT)$,  with an attempt frequency ($\bar{\nu}$) equal to the OH stretching frequency ($\nu=3700$ 1/cm), one finds a dissociation rate of one molecule out of four every $\sim 300$ s. 
In addition, if hydroxyl groups are present in the precursor water layer, they will most likely decorate the steps rather than the terraces and give rise of partially dissociated wires already upon adsorption and successive desorption from the terraces~\cite{Picolin:2009p3988}. 
 
\subsection{Nuclear quantum effects}

Effects due to the quantum nature of the nuclei play a major role in determining the structure of water adsorbed on metal surfaces~\cite{Li:2010p4068} and the strength of HBs~\cite{Walker:2010p4069}. The main effect of quantum delocalization of protons on binding and dissociation energies comes from the zero point energy (ZPE) of the vibrational modes, which can induce significant shift in both energetics and kinetics of dissociative adsorption~\cite{Grabow:2008p3906,German:2010p3674}. 
ZPE has been shown to decrease the binding energies of water at metal surfaces, since free translational and rotational modes are turned into vibrations~\cite{Feibelman:2003}.
On the other hand the adsorption and the formation of H-bonds  between adsorbed molecules induce a softening of the OH stretching modes, providing a ZPE contribution, which favors adsorption. This effect may be enhanced when adsorption occurs at step edges, since the metal-oxygen bond is stronger.\cite{Murakhtina:2006p4155}
For undissociated water dimer, trimer and periodic chains we find that ZPE favors adsorption, augmenting E$_{ads}$ by 0.08 eV/H$_2$O, whereas it reduces E$_{ads}$ for the monomer by 0.05 eV.

Hereafter we focus on the effect of ZPE on the energetics of dissociation.
The ZPE correction to dissociation energies is expressed as:
$\Delta E_{ZPE}= \sum_i h\Delta \nu_i/2$, where the sum runs over the vibrational modes of the adsorbed species and $\Delta \nu_i$ is the difference between the frequencies of intact and dissociated adsorbates~\footnote{The contribution from the vibrations of surface atom proved negligible ($<0.01$ eV) in the calculation for the monomer and was not considered in the other cases.}
The resulting ZPE corrections to E$_{diss}$ are listed in Tab.~\ref{Tab:energy}. The corrections are of the order of 0.1 to 0.2 eV in favor of the dissociated system. The main reason is that upon dissociation the contribution to ZPE from a covalent OH bond ($h\nu/2 \sim 0.23$ eV) is lost and replaced by the lower contribution from the stretching mode of the proton adsorbed on Pt ($h\nu/2 \sim 0.06$ eV). 
The other vibrational modes in the water clusters readjust after dissociation to give slightly different values for $\Delta E_{ZPE}$. 
The strength of the chemical O-Pt bonds and the presence of HBs determine the frequency of the modes of partially dissociated clusters and chain. 
In particular, the strength of the HB between H$_2$O and OH$^-$ is responsible for a significant softening of the stretching modes involved therein. For example in the case of the dimer the stretching frequency of the OH bond involved in the HB is reduced from $\sim 3000$ to $\sim 1667$ cm$^{-1}$  upon dissociation. The softening of this mode is milder for larger clusters, because the hydroxyl is clamped between two intact molecules, both donating one HB, so that each HB is weaker than the one formed in the dissociated dimer.
The bending frequencies of intact H$_2$O molecules are affected by the stronger nature of water-hydroxyl HBs in dimers, but they do not vary significantly in larger clusters. 

\subsection{Electronic structure} 

Water dissociation strengthens the oxygen-platinum interaction, and this affects the molecular levels of water as well as the local electronic structure of the edge atoms of the platinum step. We probe these effects by projecting the electronic density on atomic orbitals, namely the {\sl p}-states of oxygen and the {\sl d}-states of platinum. The projected densities of states (pDOS) show (see Supplementary Information) that the oxygen {\sl p}-states of intact water lie several eV below the Fermi level and are well localized in energy. When water dissociates, the {\sl p}-states hybridizes with the {\sl d}-band of platinum and broadens, yielding a significant density at the Fermi level. This major change in the molecular levels of water occurs upon both dissociation of  isolated molecules and partial dissociation of clusters. In the latter case, also the molecular levels of the remaining intact water molecules undergo significant broadening and hybridization with the {\sl d}-states. 

\begin{figure}[tbp]
\centerline{\includegraphics[width=8.4cm]{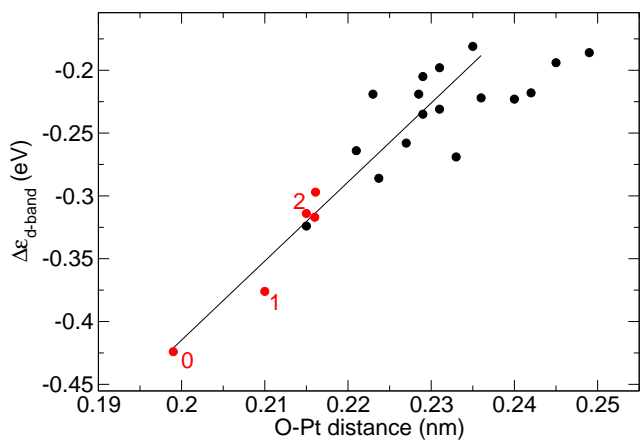}}
\caption{Shift of the barycenter of the {\sl d}-band of Pt atoms at the edge of the step as a function of the O-Pt distance, for H$_2$O (black) and $OH^-$ (red) species. The number of hydrogen bonds accepted by the hydroxy ion is indicated in the figure. The maximum number of accepted hydrogen bonds is two, hence the presence of three points with label '2', relates to the clusters or chains of length three, four, and six.}
\label{Fig:dband}
\end{figure}

According to the ``{\sl d}-band model''
the interaction with adsorbates shifts the barycenter of the {\sl d} band of metals with respect to the Fermi level by an energy $\Delta \varepsilon_d$, which is proportional to the strength of the interaction~\cite{AbildPedersen:2007p3904,Norskov:2011p3811}. 
We evaluated $\Delta \varepsilon_d$ for the edge atoms of Pt in the presence of intact and partially dissociated water clusters/chains.  
Fig.~\ref{Fig:dband} shows that $\Delta \varepsilon_d$ correlates very well with the O-Pt distance, which is an indirect indicator of the strength of the chemical bond. Surprisingly the correlation is irrespective of the dissociation state of water: hydroxy ions (red dots) and intact water molecules (black) fall on the same trend. 
Analyzing  the clusters we observed that the shift in the {\sl d}-band is localized on the atoms in direct contact with water molecules, whereas neighboring atoms at the step do not display a relevant shift. 
 This localization is further confirmed by looking at the correlation between the shift of the {\sl d}-band and the number of adsorbed molecules (see Fig. S2 in Supplementary Information): we found that the d-band shift for platinum atoms in contact with the dissociated molecule the shift depends only on the number of H-bonds and not on the number of molecules in the cluster/chain.
Fig.~\ref{Fig:dband} also indicates that the strength of the O-Pt chemical bonds and $\Delta \varepsilon_d$ decreases substantially as a function of the number of HBs accepted by the adsorbed hydroxyl ion, showing once more that the chemistry of the system is ruled by the strong interplay between chemical bonding and hydrogen bonding. 

\section{Conclusions}

Our results confirm that water on Pt binds preferentially to the edge of the steps than to (111) terraces, and suggest that water adsorption is partially dissociative. Partial dissociation occurs only when the coverage is sufficient to produce trimers or larger clusters, and is promoted by the combination of oxygen-Pt interaction and hydrogen bonding. We stress that either hydrogen-bonding or interaction of single water molecules with stepped Pt surfaces alone would not suffice to favor dissociation.
Dissociation barriers of clusters and periodic chains are sufficiently low to make dissociation possible over experimental time scales. 
We predict that, when periodic chains are formed and isolated at the Pt(221) step edge, one out of three/four water molecules dissociates into OH and H, and the proton is adsorbed at the terraces. 

Spontaneous partial dissociation of water wires makes them viable ionic conductors, since the barriers for proton hopping along hydrogen-bonded chains were shown to be relatively low, at most 0.15 eV~\cite{Scipioni:2011p3934}. Conduction would occur {\it via} diffusion of proton vacancies (holes).  Our calculations predict a high concentration of such carriers at equilibrium, but the dissociation rate could be reduced by co-adsorbing hydrogen or any other molecule that inhibits the adsorption of the dissociated protons at the {\sl fcc} hollow sites on the Pt(111) terraces, or it may be enhanced by co-adsorbing oxygen.

These results have also profound implications for catalysis and photoelectrochemical cells, where platinum surfaces are commonly used as co-catalysts or electrodes, since water dissociation at the steps would significantly shift the phase diagrams for electrochemical dissociation of water and modify the kinetics of the dissociation reaction.  

\begin{table*}[bth]
\caption{Adsorption and dissociation energies  of water clusters and periodic chains at the step edge of Pt(221). $\varepsilon_{diss}$ is the dissociation energy per water molecule. Zero point energy corrections to dissociation energies and the corrected values are also reported. }
\begin{center}
\begin{tabular}{l|cccccc}
\hline\hline 
configuration & E$_{adsorption}$ & E$_{dissociation}$  & $\varepsilon_{diss}$ & \  $\Delta E_{ZPE}$ \ & E$_{diss}+ \Delta E_{ZPE}$ & $\varepsilon_{diss}$+ $\Delta E_{ZPE}/N$ \\ 
		  &  (eV/H$_2$O)       & (eV) 	& (eV/H$_2$O)	&   (eV) 		& (eV)  	& (eV/H$_2$O)\\
\hline 
 monomer       	& 0.48 		   &  0.51 	& 0.51		& -0.18 	& 0.33   	& 0.33 \\
 dimer          & 0.55 		   &  0.20 	& 0.10		& -0.20		&  0.00  	& 0.00  \\ 	
 trimer         & 0.57 		   &  0.01 	& 0.003		& -0.13		&  -0.12 	& -0.04\\
 chain (2H$_2$O) & 0.59 	   &  0.17 	& 0.085		& -0.18  	& -0.01 	& -0.005 \\
 chain (4H$_2$O) & 0.59 	   & -0.10 	& -0.025	& -0.12 	& -0.22  	& -0.055\\
 chain (6H$_2$O) & 0.59            & -0.08	& -0.013		& -0.12         & -0.20		& -0.033 \\
 chain (12H$_2$O) & 0.59   & -0.08 & -0.0067 & - & - & - \\
\hline\hline
\end{tabular}
\end{center}
\label{Tab:energy}
\end{table*}

\section{Acknowledgements}
We thank Matthias Scheffler, Rengin Pek\"oz and Ellen H. G. Backus for discussions and critical reading of the manuscript.

\providecommand*\mcitethebibliography{\thebibliography}
\csname @ifundefined\endcsname{endmcitethebibliography}
  {\let\endmcitethebibliography\endthebibliography}{}

\end{document}